\documentclass[prl,epsf,showpacs,twocolumn,superscriptaddress,]{revtex4-1}
\usepackage[pdftex]{graphicx}
\usepackage{dcolumn}
\usepackage{bm}
\usepackage{epsfig}
\usepackage{latexsym}
\usepackage{amsmath}
\usepackage{amsfonts}
\usepackage{color}
\usepackage{array}
\usepackage{braket}
\usepackage{bbold}
\usepackage{dsfont}
\usepackage{mathrsfs}
\usepackage{tikz}
\usetikzlibrary{quantikz}
\usepackage{hyperref}
\hypersetup{
    colorlinks,
    citecolor=blue,
    filecolor=red,
    linkcolor=magenta,
    urlcolor=red
}

\makeatletter
\DeclareFontFamily{OMX}{MnSymbolE}{}
\DeclareSymbolFont{MnLargeSymbols}{OMX}{MnSymbolE}{m}{n}
\SetSymbolFont{MnLargeSymbols}{bold}{OMX}{MnSymbolE}{b}{n}
\DeclareFontShape{OMX}{MnSymbolE}{m}{n}{
    <-6>  MnSymbolE5
   <6-7>  MnSymbolE6
   <7-8>  MnSymbolE7
   <8-9>  MnSymbolE8
   <9-10> MnSymbolE9
  <10-12> MnSymbolE10
  <12->   MnSymbolE12
}{}
\DeclareFontShape{OMX}{MnSymbolE}{b}{n}{
    <-6>  MnSymbolE-Bold5
   <6-7>  MnSymbolE-Bold6
   <7-8>  MnSymbolE-Bold7
   <8-9>  MnSymbolE-Bold8
   <9-10> MnSymbolE-Bold9
  <10-12> MnSymbolE-Bold10
  <12->   MnSymbolE-Bold12
}{}

\let\llangle\@undefined
\let\rrangle\@undefined
\DeclareMathDelimiter{\llangle}{\mathopen}%
                     {MnLargeSymbols}{'164}{MnLargeSymbols}{'164}
\DeclareMathDelimiter{\rrangle}{\mathclose}%
                     {MnLargeSymbols}{'171}{MnLargeSymbols}{'171}

\newcommand{\beginsupplement}{%
        \setcounter{table}{0}
        \renewcommand{\thetable}{S\arabic{table}}%
        \setcounter{equation}{0}
        \renewcommand{\theequation}{S\arabic{equation}}%
        \setcounter{figure}{0}
        \renewcommand{\thefigure}{S\arabic{figure}}%
     }

\newcommand{\fref}[1]{Fig.~\ref{#1}}

\newcommand{\diagram}[1]{\vcenter{\hbox{\includegraphics[scale=0.45]{./#1.pdf}}}}

\newcommand{\diagramD}[1]{\vcenter{\hbox{\includegraphics[scale=0.35]{./#1.pdf}}}}

\newcommand{\DenseU}{\includegraphics[scale=0.005,height=1em]{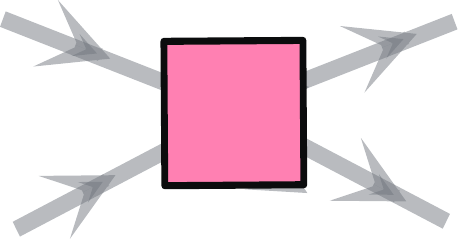}}
\newcommand{\DenseT}{\includegraphics[scale=0.005,height=1em]{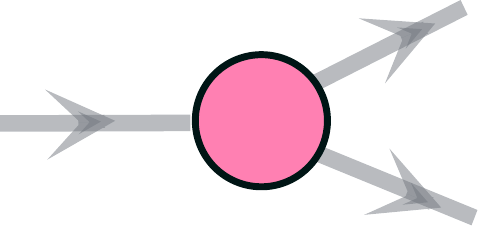}}

\newcommand{\TQgate}{\includegraphics[scale=0.005,height=1em]{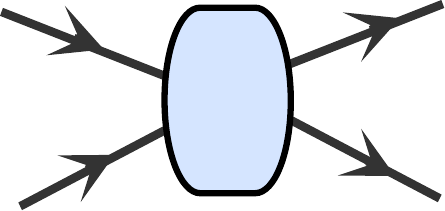}}
\newcommand{\injectqubit}{\includegraphics[scale=0.005,height=1em]{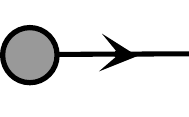}}

\newcommand{\measurment}{\includegraphics[scale=0.005,height=1em]{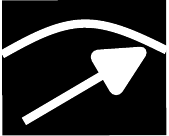}}

\newcommand{\measurmenttwo}{\includegraphics[scale=0.005,height=1em]{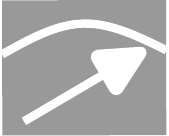}}

\newcommand{\rxx}{\includegraphics[scale=0.005,height=2.2em]{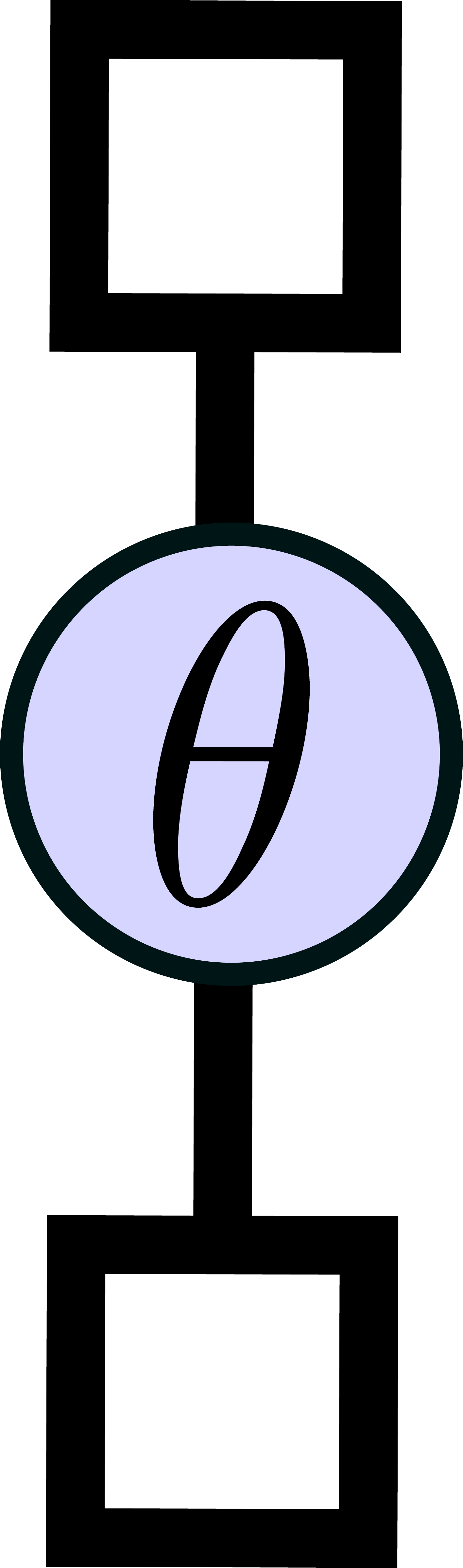}}
\newcommand{\ryy}{\includegraphics[scale=0.005,height=2.2em]{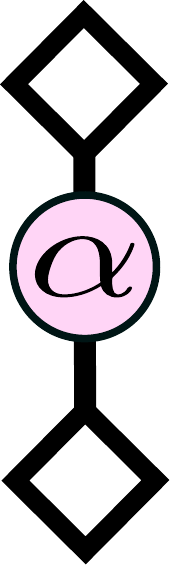}}
\newcommand{\rz}{\includegraphics[scale=0.005,height=1em]{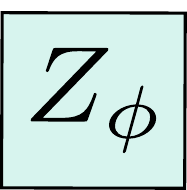}}

\begin{document}

\title{Probing critical states of matter on a digital quantum computer}

\author{Reza Haghshenas}
\email{reza.haghshenas@quantinuum.com}
\author{Eli Chertkov}
\author{Matthew DeCross}
\author{Thomas M. Gatterman}
\author{Justin A. Gerber}
\author{Kevin Gilmore}
\author{Dan Gresh}
\author{Nathan Hewitt}
\author{Chandler V. Horst}
\author{Mitchell Matheny}
\author{Tanner Mengle}
\author{Brian Neyenhuis}
\author{David Hayes}
\author{Michael Foss-Feig}
\email{michael.feig@quantinuum.com}
\affiliation{Quantinuum, 303 S. Technology Ct., Broomfield, Colorado 80021, USA}

\begin{abstract}
Although quantum mechanics underpins the microscopic behavior of all materials, its effects are often obscured at the macroscopic level by thermal fluctuations.  A notable exception is a zero-temperature phase transition, where scaling laws emerge entirely due to quantum correlations over a diverging length scale. The accurate description of such transitions is challenging for classical simulation methods of quantum systems, and is a natural application space for quantum simulation. These quantum simulations are, however, not without their own challenges \textemdash~representing quantum critical states on a quantum computer requires encoding entanglement of a large number of degrees of freedom, placing strict demands on the coherence and fidelity of the computer's operations. Using Quantinuum's H1-1 quantum computer, we address these challenges by employing hierarchical quantum tensor-network techniques, creating the ground state of the critical transverse-field Ising chain on 128-sites with sufficient fidelity to extract accurate critical properties of the model. Our results suggest a viable path to quantum-assisted tensor network contraction beyond the limits of classical methods.

\end{abstract}
\maketitle

Simulating quantum systems is a natural task for quantum computers, and is widely considered amongst their most important and most feasible near-term applications. Despite this consensus, it would be wrong to infer that the path to outperforming the best classical methods for simulating quantum systems is easy, or even clearly laid out.  One challenge is that quantum states with low entanglement can be accurately and efficiently represented using classical tensor-network (TN) techniques \cite{orus_review_2019}, i.e., low-entanglement problems are not hard for classical computers.  On the other hand, quantum states with high entanglement are generally difficult to accurately produce on existing quantum processors due to hardware imperfections. An important situation in which classical tensor network methods reveal their vulnerability to entanglement is the study of quantum critical points.  In one spatial dimension, matrix product states (MPS) cannot accurately describe critical systems in the large system-size limit unless the bond-dimension (and therefore classical simulation overhead) grows polynomially with system size \cite{orus2014}.  In practice, the polynomial overheads are tame enough to allow accurate MPS calculations for 1D critical systems, though with considerably more difficulty than calculations away from criticality. In dimensions $d>1$ or for systems out of equilibrium, the growth of entanglement with either system size or evolution time remains a significant obstacle to performing accurate TN calculations classically.

In this paper, we show that with the combination of (1) state-of-the-art classical algorithms for representing correlated quantum states \cite{PhysRevLett.99.220405,PhysRevLett.101.110501} and (2) quantum computers capable of exploiting those algorithms \cite{pino2020}, it is already possible to accurately represent critical states of matter and quantitatively extract their critical properties. In particular, we start with the multi-scale entanglement renormalization ansatz (MERA) \cite{PhysRevLett.101.110501} with a general bond-dimension $\chi$, and embed all of its tensors into parameterized unitary circuits acting on qubits. By variational minimization of the energy on a classical computer, we find a candidate circuit that can accurately capture the correct critical decay of correlations for the 1D transverse-field Ising model on a 128-site lattice.  Through a combination of circuit compression via qubit reuse \cite{decross2022qubit,fossfeig2020,Barratt2021,Chertkov2022,Niu2021,Zhang2022,chertkov_fqcp_2022}, symmetry-based error heralding made possible by the MERA structure, and zero-noise extrapolation \cite{PhysRevLett.119.180509}, we measure the decay of ferromagnetic correlations at the critical point on the Quantinuum H1-1 trapped-ion quantum computer \cite{Chertkov2022,ryan2022implementing} using 20 qubits, obtaining an estimate of $\eta=0.26 \pm 0.02$ (in good agreement with the exact value of $\eta=1/4$) and a ground state energy of $-1.27\pm 0.01$ (in good agreement with the exact value of $-4/\pi\approx -1.273$).  While this procedure presumes the ability to contract the MERA classically (since we do not do the variational optimization on the quantum computer), and does not constitute a fully quantum algorithm for finding a critical ground state, our aim in this paper is more modest but nevertheless important: We wish to show that such tensor networks lead to quantum circuits with sufficiently low resources to be feasibly and accurately executed on existing quantum computers.  The capabilities demonstrated in this paper can thus be viewed as a necessary stepping stone toward the ultimate goal of studying less classically tractable quantum behavior using quantum-assisted tensor-network calculations, such as studying near-critical dynamics (see discussion at the end of the manuscript), or exploring the expressive power of classically intractable MERA generalizations \cite{kim2017robust}.

\begin{figure*}[!t]
\includegraphics[width=2.0\columnwidth]{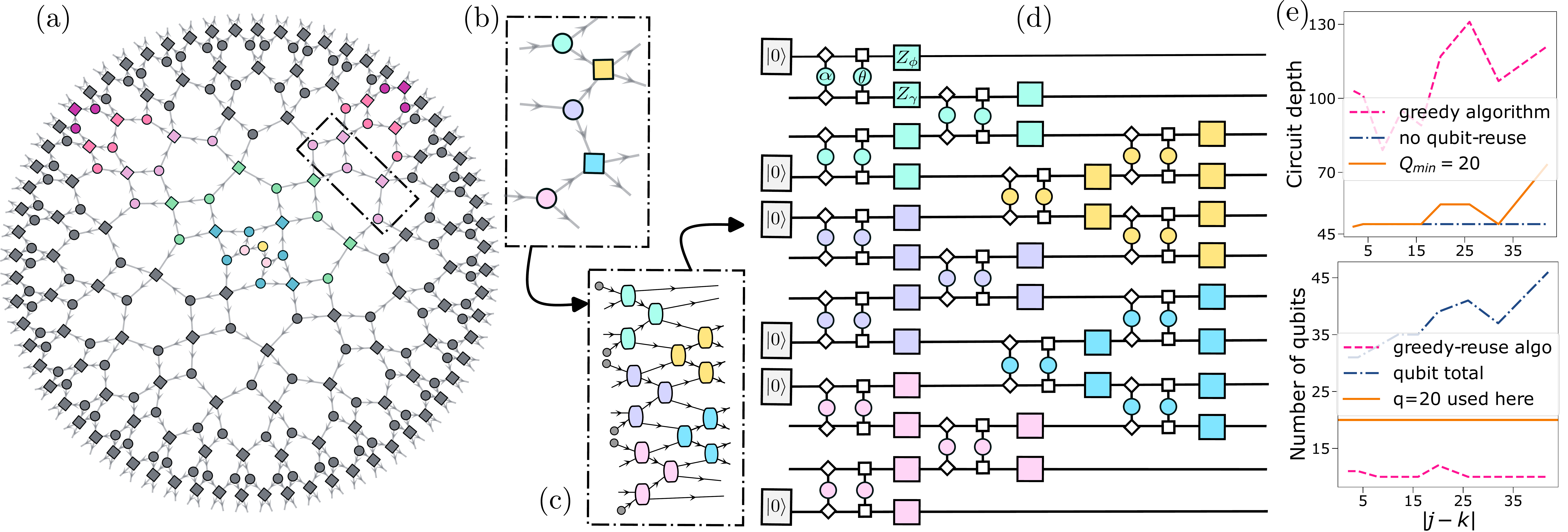}
\caption{MERA on a quantum computer. 
 (a) Abstract representation of the 128-site MERA used in this work.  Circles represent isometries while squares represent disentanglers, and the arrows flow in the direction of circuit time.  We use a bond dimension $\chi=4$, meaning that each line in (a) carries two qubits and each isometry ingests two initialized qubits [shown in (c,d) but not (a,b)].  (b) Detail of a block of disentanglers and isometries, which are decomposed into circuits consisting of two-qubit gates using a brickwall ansatz (c). (d) Individual two-qubit gates are parameterized in terms of two native parametrized entangling gates [see below Eq.\,(\ref{EQ:Tensor-gate}) for the gate definitions] in a fashion that respects a global $\mathbb{Z}_2$ symmetry. Parameters are only shown for one such two-qubit gate, but each has independent parameters for a total of around $2000$ independent parameters in the ansatz. (e) Lower panel shows qubit resources with maximal greedy qubit reuse, without any reuse, and with reuse that fully exploits the 20 qubits currently available on the Quantinuum H1-1 quantum computer.  Upper panel shows the circuit depth \cite{tket2021} for the same three choices.}
\label{fig:MERA_full}
\end{figure*}

\emph{Sampling a MERA on a quantum computer.}---Of all tensor network methods for the classical simulation of quantum systems, MERA is the most naturally interpreted as a quantum circuit: while its name highlights the action of MERA as a coarse-graining transformation that respects local entanglement structure \cite{PhysRevLett.99.220405}, from its inception MERA was also envisioned in the reverse direction as a quantum circuit in which all isometries are embedded in unitaries \cite{PhysRevLett.101.110501}.  In this interpretation, an $L$-site MERA is a circuit ansatz that takes as input $L$ ancilla qubits initialized to the $\ket{0}$ state and produces as output an entangled state.  As a circuit ansatz, MERA has a number of valuable properties.  On the physical level, the MERA circuit is specifically designed to capture scale invariant quantum states, e.g. ground states at continuous quantum phase transitions.  On a more technical level, the causal structure of the ansatz allows for relatively efficient tensor contractions in order to compute the expectation value of operators supported on a finite number of sites (with run time scaling polynomially with $\chi$ and logarithmically with $L$).

When MERA is viewed as a quantum circuit, these contraction efficiencies equate to reductions in the required quantum resources, e.g., the number of qubits and gates required to sample a local region of the output state. More specifically, the output state over a region containing a finite number of sites (e.g., some small set of sites in order to compute a correlation function) has a past causal cone containing $O(\log L)$ gates. Since a constant fraction of the gates are isometries---each of which ingests a constant number of initialized qubits--- that output state can always be sampled using a circuit acting on $O(\log L)$ qubits. If qubits are allowed to be reset and reused, the quantum cost for sampling a local observable is even lower \cite{miao2021quantum, Miao:2023Convergence, anand2022holographic}. The constant size of the causal cone implies that qubits must exit the cone as fast as new qubits are introduced through isometries; if not, the causal cone would grow with time. Resetting qubits upon exit for use in future isometries must therfore allow the output state to be reached with a constant ($L$-independent) number of qubits. The non-unitary channel resulting from resetting qubits upon exit and reusing them in pending isometries is essentially a physical realization of the descending MERA superoperator \cite{PhysRevB.79.144108}.  While we do not explicitly explore this feature here, it is even possible to sample the full MERA output (and therefore estimate arbitrary non-local observables) using only $O(\log L)$ qubits. In this latter case,
despite the larger number [$O(L)$] of gates in the circuit, each individual output site still has only $O(\log L)$ gates that causally impact it, greatly limiting the damage of gate errors to the measurement outcomes.

\emph{MERA for the TFIM at criticality.}---The transverse-field Ising model (TFIM) in one dimension and with periodic boundary conditions is described by the Hamiltonian
\begin{align}
H=-J\sum_{j=1}^{N}X_j X_{j+1}-h\sum_{j=1}^{N}Z_j,
\end{align}
where $X$ and $Z$ are Pauli matrices and we take the subscripts modulo $N$ to give periodic boundary conditions.  The TFIM (with long-range interactions) has been extensively studied in analog trapped-ion quantum simulators because its interactions are natural to implement \cite{islam_2011,britton_2012}.  We note that in our digital approach, the specific form of the two-body interactions is of little consequence, and we primarily chose the TFIM for its conceptual simplicity.  The TFIM undergoes a continuous phase transition from a paramagnet to a ferromagnet as $h\rightarrow J$ from above, with the critical point at $h=J$ in the universality class of the $2D$ classical Ising model. Universal properties of the model manifest in a variety of equal-time correlation functions in the ground state, but we focus on the correlator describing fluctuations of the order parameter, $\mathcal{C}_{xx}(r)= \langle X_j X_{j+r}\rangle-\langle X_j\rangle \langle X_{j+r}\rangle$, which decays asymptotically as $\mathcal{C}_{xx}(r)\sim 1/r^{\eta}$ with $\eta=1/4$.



We minimize the energy $\langle H \rangle$ using a bond-dimension $\chi= 4$ binary MERA with a single top-tensor, depicted in \fref{fig:MERA_full}(a) \footnote{In an effort to minimize gate counts, we explored various schemes interpolating between MERA and a tree tensor network \cite{shi2006} (e.g. excluding various layers of disentanglers), and found that removing the top layer of disentanglers had only minimal effect on the convergence of the energy; we therefore excluded that one row of (two) disentanglers.}.  A MERA is built out of coarse-graining isometry tensors $\DenseT$ and so-called disentangler tensors $\DenseU$. Both tensors (with $\chi= 4$) can be decomposed into a local brick-wall circuit acting on 4 qubits, which we depict graphically as
\begin{align} 
\label{EQ:DenseTU-local}
\diagramD{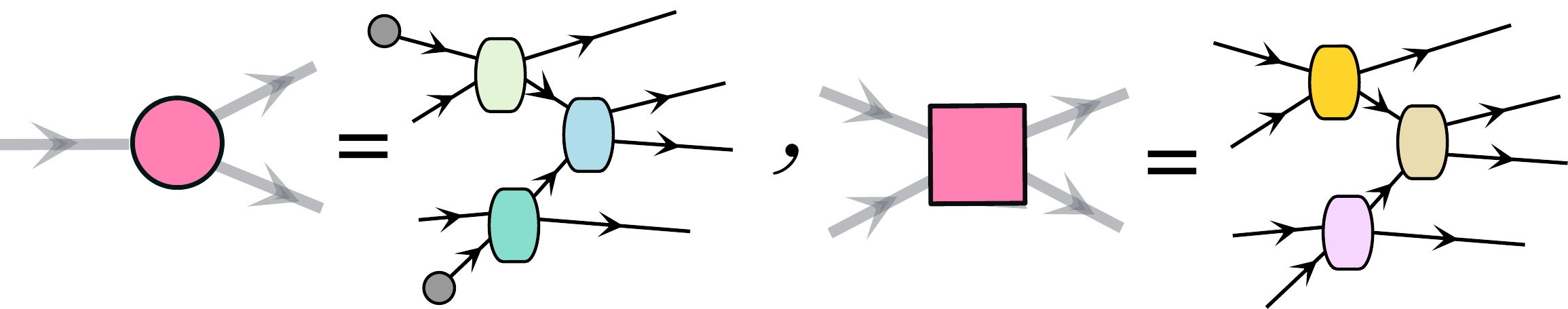}
\end{align}
The restriction of the brick-wall circuit to depth 2 reduces the number of variational parameters in the MERA \cite{RH2022} (relative to maintaining arbitrary tensors at bond dimension $\chi=4$), though we find that a depth-2 brickwall ansatz is quite effective at producing low energies and accurate correlation functions.  Here, $\injectqubit$ stands for injecting a qubit initialized to $\ket{0}$, and $\TQgate$ represents a two-qubit unitary gate. The latter are further decomposed as
\begin{align}
\label{EQ:Tensor-gate}
\diagram{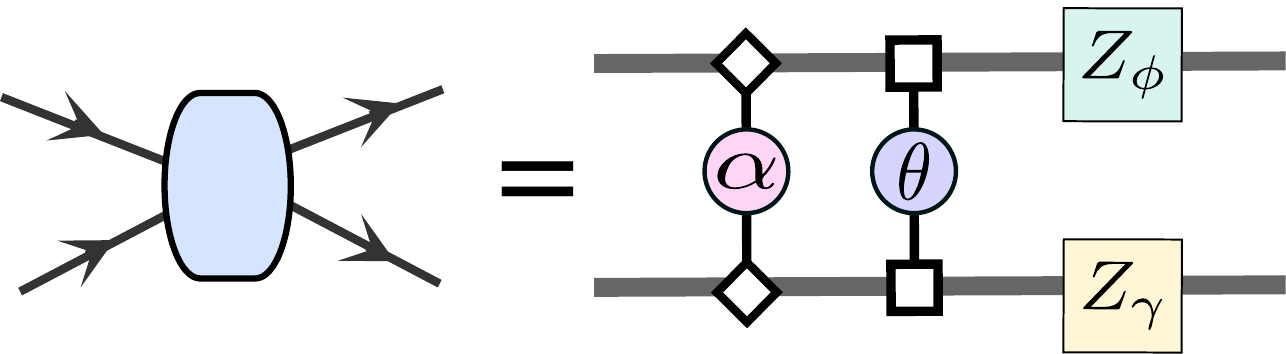}
\end{align}
where gates $\rxx$ and $\ryy$ stand for $U_{XX}(\theta)=\exp(-i\frac{\theta}{2} X\otimes X)$ and $U_{YY}(\alpha)=\exp(-i\frac{\alpha}{2} Y\otimes Y)$. Up to single-qubit rotations, they are equivalent to the parameterized entangling gate $U_{ZZ}(\theta)=\exp(-i\frac{\theta}{2} Z\otimes Z)$ native to the Quantinuum H1 quantum computer. The symbol $\rz$ represents a single-qubit $Z$-axis rotation gate $R_Z(\phi) = \exp(-i\frac{\phi}{2} Z)$. This decomposition is chosen to respect the $\mathbb{Z}_2$ symmetry of the TFIM, and ensures that the output of the MERA is in the same symmetry sector as the ground state. The resulting quantum circuit is referred to as a quantum circuit MERA (qMERA).  

The $\mathbb{Z}_2$-symmetric qMERA is optimized by using a global gradient-based method, with derivatives analytically calculated by automatic differentiation \cite{gray2018quimb, RH2021, RH2022}. The ansatz provides a relative energy error $\delta \langle H \rangle \sim 10^{-5}$ and produces two-point correlation functions with relative error $\delta \mathcal{C}_{xx}(r) \sim 10^{-2}$, compared to the exact results, for all distances. We note that the distribution of optimized rotation angles is strongly peaked around zero (see Supplemental Material). This feature of the optimized ansatz is extremely desirable from a fidelity standpoint, as the fidelity of the native parameterized entangling gates improves with decreasing angle \footnote{Extensive theoretical modeling and experimental evidence from direct randomized benchmarking indicates that the error (average infidelity) of our native gates decreases roughly linearly with angle (with an offset at zero angle of well under $10^{-3}$).}.   

\begin{figure}[!t]
\includegraphics[width=0.85\columnwidth]{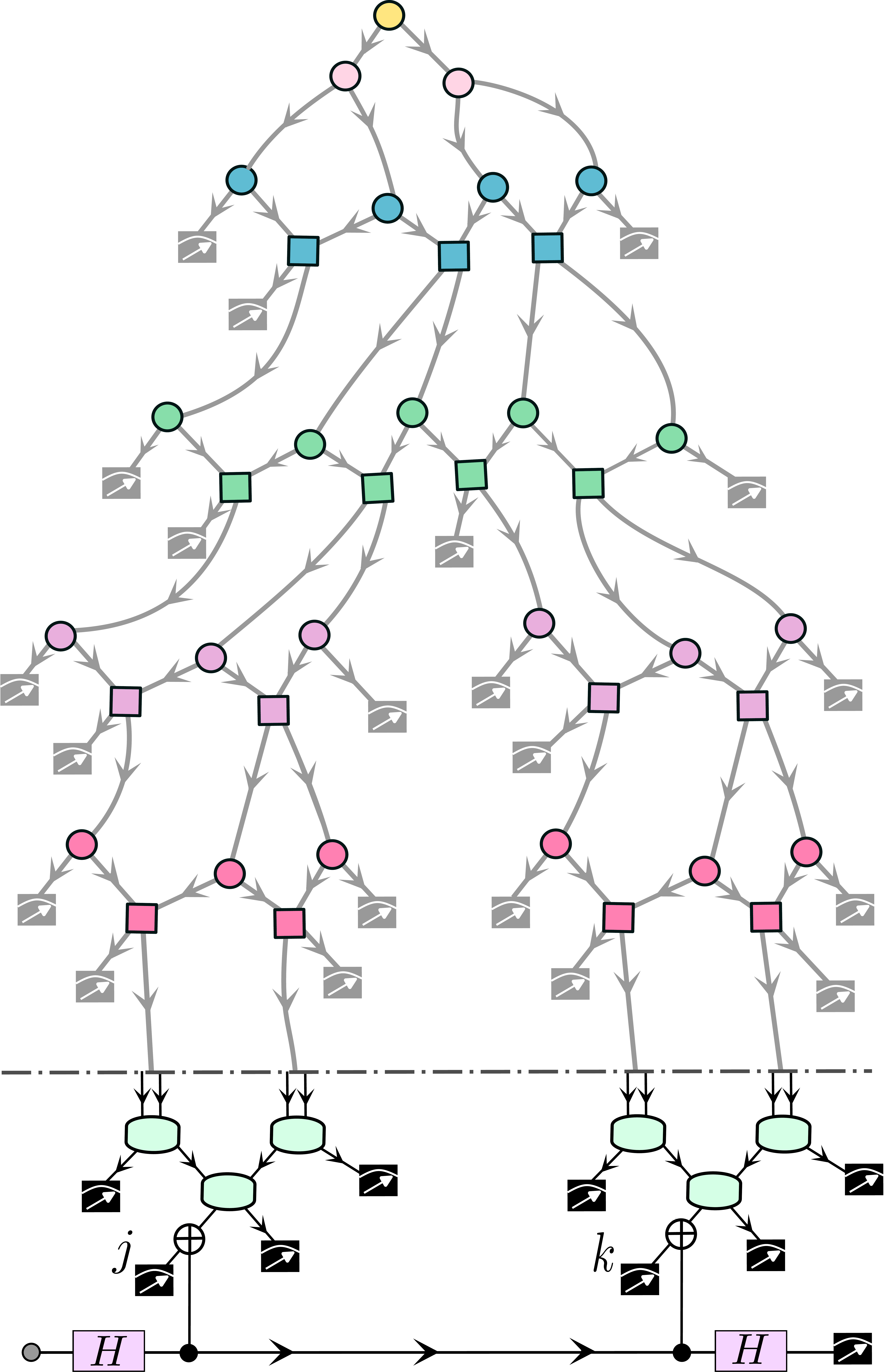}
\caption{Measuring qMERA correlation functions with symmetry-based error mitigation. The full past-causal cone of the two output sites $j$ and $k$ is shown.  All qubits exiting the past causal cone are measured in the $Z$ basis, while we measure both $X_jX_k$ (using an ancilla qubit) and $Z_jZ_k$ in order to compute the correlation function $\langle X_jX_k\rangle$ with data post-selected on the absence of parity-violations. The symbols $\measurmenttwo$ and $\measurment$ stand for two-qubit and one-qubit measurements in the computational basis, respectively.  }
\label{fig:MERA_corr}
\end{figure}

When computing correlation functions between sites $j$ and $k$ on the quantum computer, the restriction of the MERA to the past causal cone of those qubits, denoted $\mathcal{C}_{j,k}$, results in a circuit in which many qubits exit $\mathcal{C}_{j,k}$ at intermediate times and remain idle for the rest of the circuit. We therefore have many opportunities to reset such qubits and reinject them to implement isometries at later times in the circuit.  We generate lower-width circuits utilizing qubit reuse with an automated qubit-reuse compilation strategy \cite{decross2022qubit}.  Figure \ref{fig:MERA_full}(e) shows the dependence of circuit depth \footnote{Circuit depth is given by pytket-quantinuum library \cite{tket2021}, defined as interior vertices on the longest path through the directed Acyclic graph. The reported circuit depth is obtain by only keeping two-qubit gates.} and qubit number on the distance between sites $i$ and $j$ using no qubit reuse, and then using the heuristic qubit reuse compression described in Ref.\,\cite{decross2022qubit}. We note that without any qubit reuse, sampling the longest-range correlation functions we measured ($|j-k|=32$) requires $37$ qubits.  The compression algorithm also allows a specification of the minimum qubit number; setting that to 20 (to ensure that all qubits are utilized) ensures the lowest possible circuit depth given the available number of qubits.

\emph{Error mitigation.}---To mitigate the impact of gate errors on correlation function measurements we employ two separate strategies. First, we enforce the output state of the MERA to respect the global $\mathbb{Z}_2$ symmetry of the TFIM, which is generated by the product of Pauli-$Z$ operators on all output qubits, $\mathcal{Z}=\bigotimes_{j=1}^{N}Z_j$. By injecting all qubits in the $\ket{0}$ state (an eigenstate of $\mathcal{Z}$ with eigenvalue $+1$) and requiring that the component tensors all satisfy this symmetry locally \cite{PhysRevA.82.050301} (which we achieve by building them from two qubit gates that commute with $Z\otimes Z$ acting on the respective qubits), we 
 can be sure that in the absence of noise the output state is a $+1$ eigenstate of $\mathcal{Z}$.  The $\mathbb{Z}_2$ symmetry is global, and therefore any given subset of the output sites will not generally have a well-defined $Z$ parity.  However, it \emph{is} respected by any subset of the output qubits together with all qubits exiting the past causal cone of that subset, even if they have not themselves been evolved to the final layer of the MERA output.  Therefore, defining $\mathcal{C}$ to contain all qubits exiting the past causal cone of outputs $j$ and $k$ associated with a given correlation function $\langle X_j X_k\rangle$, and $\mathcal{Z}_{\rm causal}=\otimes_{a\in\mathcal{C}}Z_a$, we should have (in the absence of any errors)
 $\langle Z_j Z_k \mathcal{Z}_{\rm causal}\rangle=1$. Thus by measuring $Z_jZ_k\mathcal{Z}_{\rm causal}$, we can herald the presence of errors that change the symmetry sector of the state (which is only a strict subset of all possible errors).  Although the symmetry operator and the correlation operator to be measured ($X_jX_k$) are diagonal in different product-state bases, they commute and can be measured simultaneously.  We do so by measuring all qubits in $\mathcal{C}$ in the $Z$ basis, making an ancilla-assited parity measurement of $X_jX_k$, and then measuring out qubits $j$ and $k$ in the $Z$ basis, as in \fref{fig:MERA_corr}.  From the $Z$ basis measurements we determine whether $\langle Z_j Z_k \mathcal{Z}_{\rm causal}\rangle=1$ or not; if not we discard the associate data, and if so we keep the corresponding projective measurement of $X_iX_j$ in order to estimate the correlation function (data discard rates of $\sim 30\%$ are typical, see Supplemental Material).  In addition to this symmetry-based error heralding, we also employ a zero-noise extrapolation technique \cite{PhysRevLett.119.180509} described in the Supplemental material.

\begin{figure}[!t]
\includegraphics[width=1.0\columnwidth]{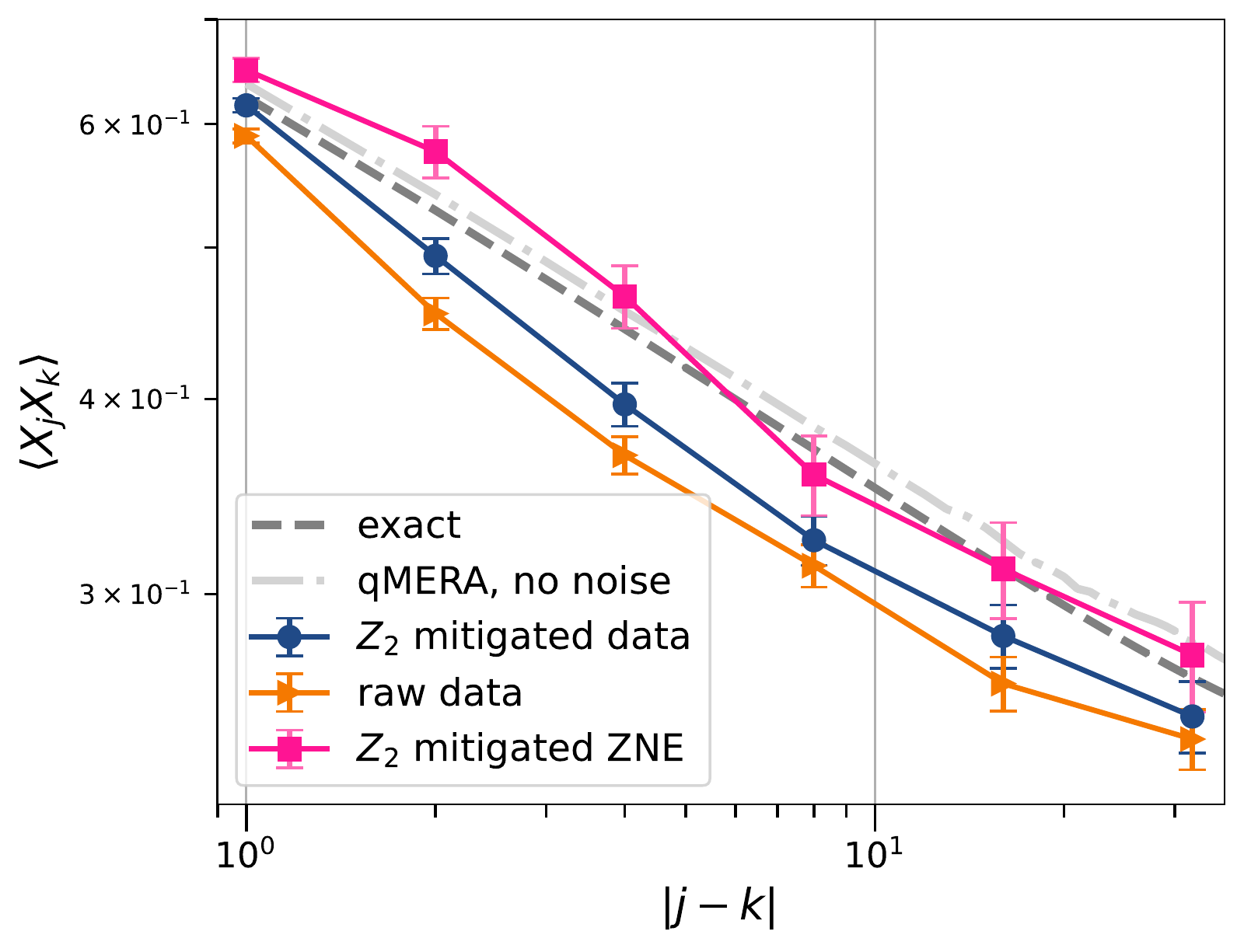}
\caption{Measured $XX$ correlation functions using qMERA. Orange points are raw data, blue points are data post-selected on absence of heralded $\mathbb{Z}_2$ symmetry violations, and pink points are obtained by applying zero-noise extrapolation (ZNE) to the symmetry heralded data.  A power-law fit to the ZNE results yields a decay exponent of $\eta_{\rm meas}=0.26\pm 0.02$, in good agreement with the exact value $\eta=1/4$ for the TFIM.}
\label{fig:results}
\end{figure}

\emph{Results.}---We implement the qMERA on Quantinuum's H1-1 quantum computer \cite{pino2020,H11,H11specsheet}, which at the time of operation was utilizing five gate zones (each operating in parallel) and $20$ qubits (each qubit is comprised of the $\ket{F=0,m_F=0}$ and $\ket{F=1,m_F=0}$ clock states of a $^{171}$Yb$^+$ ion). H1 can natively implement the parameterized entangling gate $U_{ZZ}(\theta)=\exp(-i \frac{\theta}{2}Z\otimes Z)$, and the perfect entangler $U_{ZZ}(\pi/2)$ typically operates with an average fidelity close to $99.8\%$ (gates with angles $|\theta|<\pi/2$ generally have higher average fidelity).  The qubit reuse compression required to fit the circuits onto the available 20 qubits does not change the logical structure of the circuit, and the impact of gate errors on measurement outcomes is unchanged. However, it \emph{does} impact the times at which gates are applied, and serializing gates that could in principle be implemented in parallel generally leads to longer total wall-clock time for a given circuit.  Therefore, errors associated with qubit idling (predominantly dephasing-type errors on the Quantinuum H1-1 device) will generally be made worse by qubit reuse. However, the measured memory error per qubit of $\sim 10^{-4}$ at 20ms (the typical time required for running an arbitrary set of $10$ two-qubit gates on all $20$ qubits) is small enough compared to the two-qubit gate error that this effect is tolerable for our circuits.  In Fig.\,\ref{fig:results} each distance of the correlation function is obtained from a separate experiment, and the mean value is obtained by averaging over $8000$ shots.  All error bars represent $1\sigma$ confidence intervals obtained by bootstrap resampling the data (1500 resamples). A linear fit to the post-processed data (after both symmetry-based error heralding and zero-noise extrapolation) yields a critical decay exponent of $\eta=0.26\pm0.02$ \footnote{The uncertainty in the critical exponent is reported by employing two methods of direct bootstrap sampling and finding the best fit by minimizing Chi-squared function. Both methods produce the same consistent error bar.}, in good agreement with the exact value $\eta=1/4$ (and also with the best fit to the correlation decay of the noiseless qMERA, $\eta=0.24$).

\emph{Outlook}.---Given the ability to accurately represent 1D strongly-correlated states of matter on a quantum computer, a natural generalization would be the implementation of time evolution on top of such a state, which could be used to explore a variety of near-equilibrium and non-equilibrium physics such as critical scaling of defect formation \cite{PhysRevB.106.L041109} or transport properties \cite{PhysRevB.79.214409}. Because the full output distribution of an $L$-site MERA can be sampled with $\sim \log(L)$ qubits sequentially in 1D (e.g., measuring output sites from left to right in a system with open boundary conditions), the output of a depth $d$ brick-wall circuit \emph{appended} to a MERA can be sampled with $\sim \log(L)+d$ qubits. Generalizations of MERA to effectively higher bond dimension by maintaining bond-dimension 2 but extending the depth/connectivity of the disentangler layers would also be natural for a quantum computer \cite{kim2017robust}, and yields quantum algorithms with potential exponential resource reductions in both time and memory compared to classical contraction of the corresponding network.

\begin{acknowledgments}
We thank the entire Quantinuum team for numerous contributions that enabled this work.  We are grateful to Patty Lee, Russell Stutz, Wes Campbell, and Eric Hudson for helpful comments on the manuscript. 
\end{acknowledgments}

\bibliography{refs}


\

\

\
\appendix

\section{SUPPLEMENTAL MATERIAL}
\beginsupplement

\subsection{Zero-noise extrapolation}

We employ a form of zero-noise extrapolation (ZNE) \cite{PhysRevLett.119.180509}, which works for our native arbitrary-angle two-qubit gate $U_{ZZ}(\theta)=\exp(-i\frac{\theta}{2} Z\otimes Z)$, to mitigate the errors in our experimentally measured observables. The expectation value of an observable $E(p)$ computed from the quantum computer depends on the amount of noise caused by noisy operations like two-qubit gates, with $p$ setting the scale of the noise. For small enough $p$, we expect $E(p) \approx E_0 + p E_1$ to be well approximated by a first-order Taylor expansion, with $E(p=0)=E_0$ corresponding to the ideal zero-noise expectation value. By performing two experiments, one at $p$ and one at $mp$ for small enough $m$, we can use a simple linear extrapolation to estimate the zero-noise value as 
\begin{align}
E_0 \approx \frac{m E(p) - E(mp)}{m - 1}.
\end{align}
When the error on $E(p)$ is $\sigma_1$ and on $E(mp)$ is $\sigma_m$, the approximate error on the extrapolation is
\begin{align}
\sigma_0 \approx \frac{\sqrt{m^2 \sigma_1^2 + \sigma_m^2}}{m - 1}.
\end{align}
If we suppose that $\sigma_1=a/\sqrt{N_1}$ and $\sigma_m=b/\sqrt{N_m}$, which is the case when we use $N_1$ and $N_m$ samples to estimate the standard error of $E(p)$ and $E(mp)$, then we find that 
\begin{align}
\sigma_0 = \frac{1}{m-1}\sqrt{\frac{am^2}{N_1} + \frac{b}{N_m}}.
\end{align}
The contribution of the two errors $\sigma_1,\sigma_m$ to the extrapolated error are comparable when we set $N_1 \propto m^2 N_m$. Note that here we assume that $a$ and $b$ are approximately constant, but in principle they can have non-trivial dependence on $p$ (and therefore $m$) if $p$ is large enough.

\begin{figure}[!t]
\includegraphics[width=1.0\columnwidth]{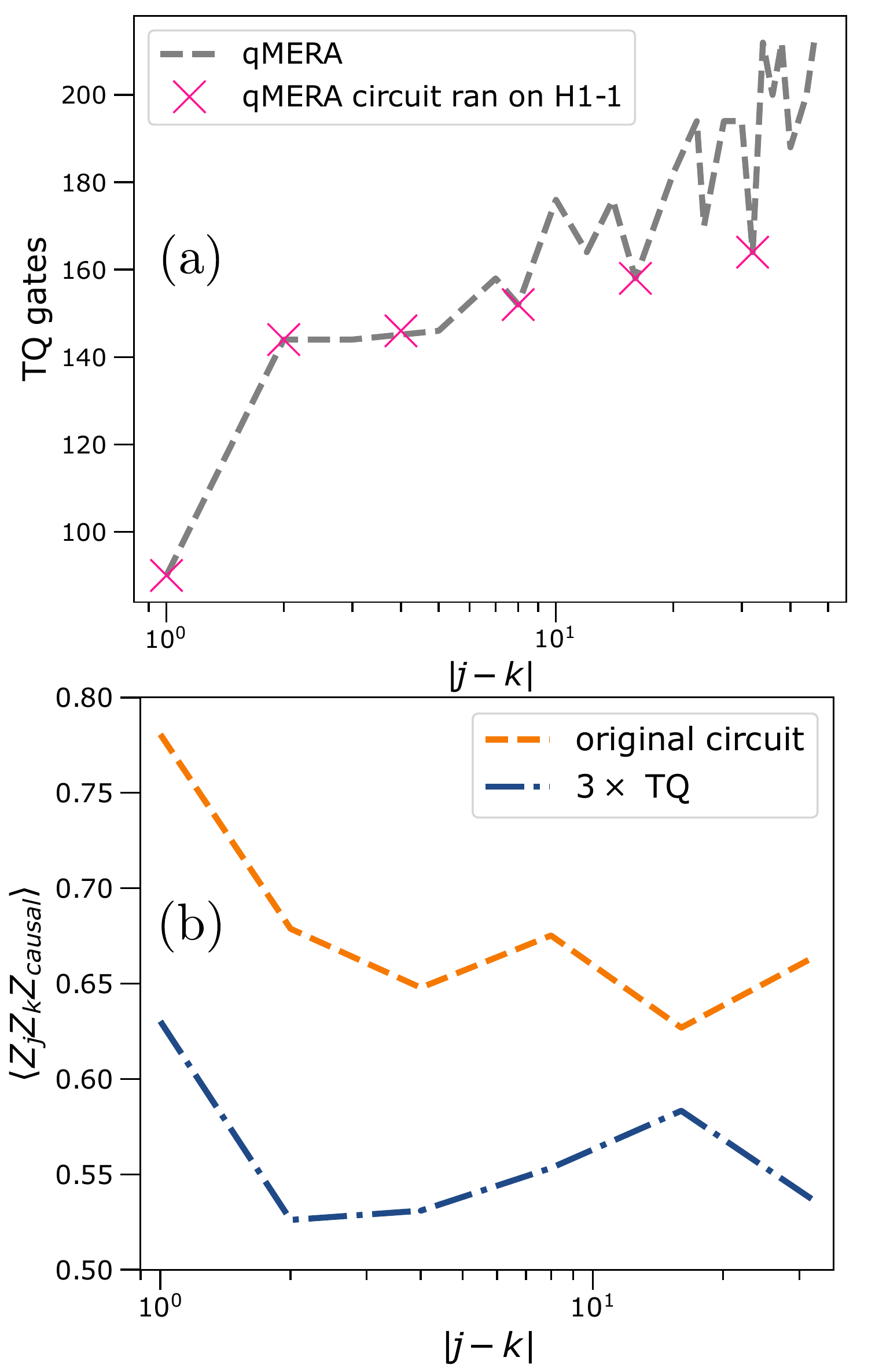}
\caption{ (a) Number of two-qubit gates versus the correlation-function distance for the qMERA ansatz used in this paper. In order to perform ZNE, a second error-amplified qMERA circuit with $3 \times$ as many two-qubit gates is required.  (b) The expectation value of the parity observable $(1 + \langle Z_j Z_k \mathcal{Z}_{\textrm{causal}}\rangle)/2$ for the original and error-amplified qMERA circuits, as measured during the calculation of $\langle X_j X_k\rangle$ correlation functions. With no errors, because of the $Z_2$ symmetry of the model, we expect $\langle Z_j Z_k \mathcal{Z}_{\textrm{causal}}\rangle=1$. The value of $(1 + \langle Z_j Z_k \mathcal{Z}_{\textrm{causal}}\rangle)/2$ corresponds to the fraction of data kept when post-selecting on data obeying the $Z_2$ symmetry. }
\label{fig:resultsCircuit}
\end{figure}

When applying ZNE, one must decide how to scale up the noise in the system in a controlled way in order to properly estimate $E(mp)$ for a fixed, known $m$. For our circuits, we scale up the noise of the two-qubit gates by a factor of $m=3$ by replacing each two-qubit gate $U_{XX}(\theta)=\exp(-i X\otimes X\theta/2)$ with an equivalent unitary that involves three copies of the gate:
\begin{align}
U_{XX}(\theta) &= U_{XX}(\theta) [U_{XX}(\theta)^\dagger U_{XX}(\theta)] \nonumber \\
&= U_{XX}(\theta) U_{XX}(-\theta) U_{XX}(\theta) \nonumber \\
&= U_{XX}(\theta) Z_1 U_{XX}(\theta) Z_1 U_{XX}(\theta) \label{eq:gatezne_equation}
\end{align}
where $Z_1 U_{XX}(\theta) Z_1 = U_{XX}(-\theta)$ because $Z_1$ and $X_1 X_2$ anti-commute. Diagrammatically, this replacement looks like
\begin{align} 
\label{EQ:gatezne}
\diagram{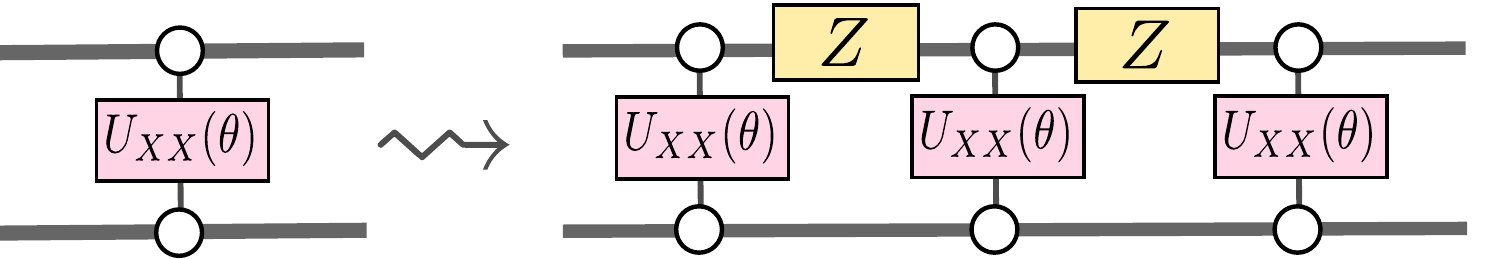}
\end{align}
Importantly, the right-hand-side of Eq.~(\ref{EQ:gatezne}) has $m=3$ copies of the same gate $U_{XX}(\theta)$ as on the left-hand-side and does not include $U_{XX}(-\theta)$ gates, since in general the error of the $\theta$ and $-\theta$ gates do not have to coincide. For our quantum computers, the two-qubit gate errors are larger than the single-qubit gate errors by about a factor of $100$, so to a good approximation the unitary on the right-hand-side has error $3p$ compared to the two-qubit gate error $p$.

\begin{figure}[!t]
\includegraphics[width=1.0\columnwidth]{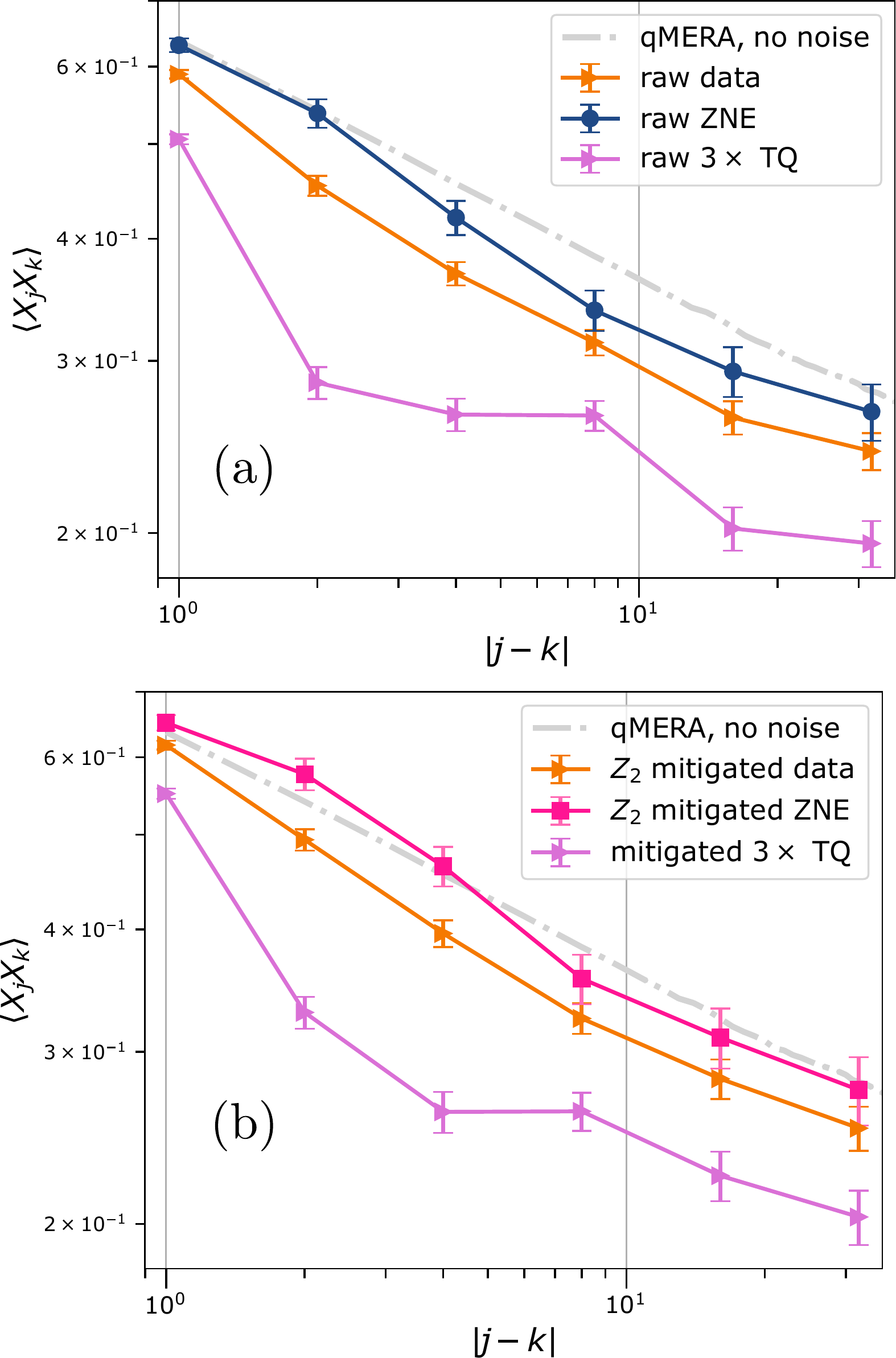}
\caption{ The experimental results from Quantinuum's H1-1 trapped-ion quantum computer. (a, b) The expectation value of the correlation function $\langle X_j X_k \rangle$ with different error mitigation strategies applied (either post-selection based on $Z_2$ symmetry, or ZNE, or both), compared to the ideal noise-less result shown as a gray dashed line. Also shown are the correlation functions for the circuits with $3\times$ as many two-qubit gates applied, which were used for ZNE.}
\label{fig:resultsZNE}
\end{figure}

The improvement in our data due to ZNE is shown in Fig.~\ref{fig:resultsCircuit} and Fig.~\ref{fig:resultsZNE}. Fig.~\ref{fig:resultsCircuit}(a) presents the number of two-qubit gates used in the qMERA experiment on Quantinuum's H1-1 trapped-ion quantum computer. In order to perform ZNE, we ran equivalent qMERA circuits with $3 \times$ as many two-qubit gates, with a total of around $\sim 480$ two-qubit gates for the deepest circuits. In Fig.~\ref{fig:resultsCircuit}(b), we have compared their performance with respect to the parity metric that measures the violation of $Z_2$ symmetry due to two-qubit gate errors. Fig.~\ref{fig:resultsZNE}(a, b) shows how ZNE combined with post-selection on having no parity violations produces quantitatively accurate experimental correlation functions consistent with ideal noiseless calculations. The circuit-level compilation was performed using the pytket-quantinuum library \cite{tket2021}, which contains compilation tools specifically optimized for Quantinuum devices.

\begin{figure*}[!t]
\includegraphics[width=2 \columnwidth]{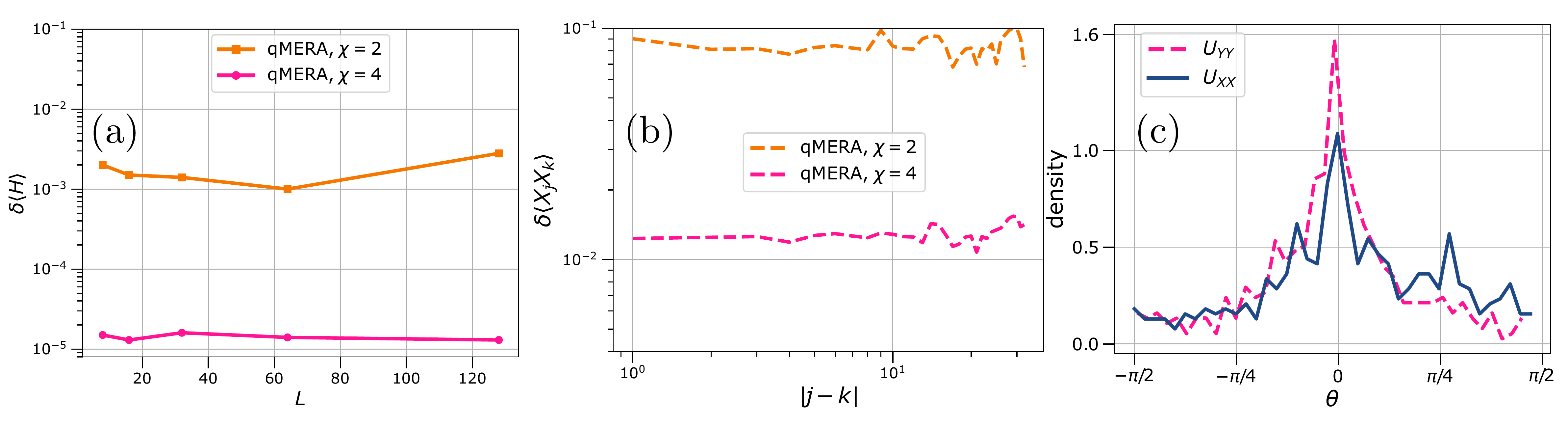}
\caption{The variational results of the qMERA ansatz for the TFIM at the critical point. (a) The relative energy error $\delta \langle H \rangle $ versus lattice size for qMERA with $\chi=\{2, 4\}$, corresponding to one and two bond qubits, respectively. We have included the reference points of dense MERA with $\chi=4$ and an MPS (obtained by DMRG algorithm) with $\chi=16$ for the largest system size $L=128$. (b) The relative error in the two-point correlation function  $\delta \mathcal{C}_{xx}(|j-k|)$ as a function of distance.  (c) Distribution of the rotational angles for gates $U_{XX}$ and $U_{YY}$ (note that the results are from the optimized values of rotational angles). The result is for the qMERA ansatz used in this paper for the experimental demonstration, i.e. lattice size $L=128$ and bond dimension $\chi=4$.}
\label{fig:MERA_classical}
\end{figure*}

\subsection{Classical optimization of qMERA}
Here, we report benchmarking of the variational capability of the qMERA ansatz. The ansatz is optimized by using a global gradient scheme, where all variational rotational angles $\Vec{\theta}$ are updated at the same time. The gradient of the energy cost function $\langle H \rangle_{\Vec{\theta}}$ is analytically calculated by automatic differentiation with respect to the rotational angles. We use optimized-order tensor-network contractions \cite{Gray2021hyperoptimized} to evaluate the cost function. The cost-function is then minimized by using the Broyden-Fletcher-Goldfarb-Shanno (L-BFGS-B) algorithm, which halts once the relative change in energy is less than $10^{-8}$. In Fig.~\ref{fig:MERA_classical}(a), we show the relative energy error (compared to the exact numerical results) versus different lattice sizes. As observed, we can achieve a consistent relative error $\sim 2 \times 10^{-5}$ for the qMERA ansatz used in this paper ($\chi = 4$). The parameterized gates (around $ \sim 2000$ parameters) are initialized by drawing their parameters from a uniform random distribution. We notice that the optimized ground-state energy does not have a strong dependence on the initial parameters, as might be explained by the absence of barren plateaus in isometric tensor networks (including qMERA) \cite{Cerver:2023, barthel:2023}. In addition, qMERA relative energy error is close to a dense-$\chi=4$ MERA (with no two-qubit gate decomposition) with relative error $\sim 1 \times 10^{-5}$, indicating that the restricted two-qubit gate decompositions have a marginal effect on the qMERA ansatz at this bond dimension. We observe that an MPS (obtained by DMRG algorithm) with $\chi=16$ provides a similar level of accuracy for the ground-state energy as qMERA.

In Fig.~\ref{fig:MERA_classical}(b), the relative error in the correlation function $\delta \mathcal{C}_{xx}(r)$ has been plotted versus different lattice distances for the qMERA, MERA and MPS on a lattice size $L=128$. The qMERA ansatz can provide a relative error $ \sim 10^{-2}$ at all distances, indicating the capability of the qMERA to capture long-range correlations.

We note that a key element distinguishing our work from Ref.\,\cite{anand2022holographic} lies in the choice of optimization ansatz. Specifically, we employ a scalable, energy-based cost function, whereas Ref.\,\cite{anand2022holographic} directly optimizes the overlap of qMERA with the exact ground-state wave function. In contrast to the latter approach, which requires a-priori knowledge of the correlation structure of the ground state, energy-based optimization (which appears to require a larger bond-dimension to achieve accurate critical exponents) paves the way toward achieving precise results in targeted observables even when the ground state wave function is not known.

Finally, we report the distribution of variationally optimized angles in the qMERA ansatz. As shown in Fig.~\ref{fig:MERA_classical}(c), it is observed that for both arbitrary-angle two-qubit gates $U_{XX}$ and $U_{YY}$ the parameter distribution is peaked near $\theta \sim 0$ and relatively narrow. This feature of the optimized ansatz is extremely useful in improving the circuit fidelity, as the error of our native gates decreases roughly linearly with angle.

\begin{figure*}[!t]
\includegraphics[width=2 \columnwidth]{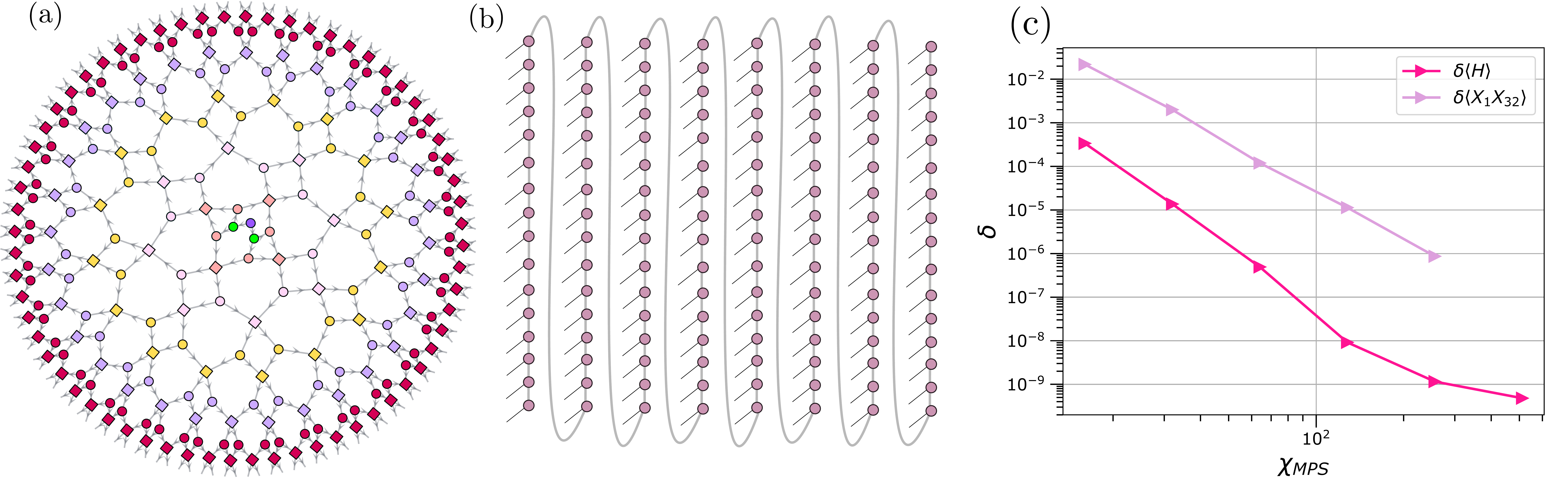}
\caption{MPS approximation of qMERA circuit used in this paper. (a) The qMERA circuit is approximated by (b) a dense MPS, defined by non-unitary tensors, with bond dimension $\chi_{MPS}$. (c) Relative error $\delta$ in the energy and long-range correlation $\langle X_{1}X_{32}\rangle$ as a function of MPS bond dimension $\chi_{MPS}$. The plot shows that long-range features of the qMERA circuit can be well approximated by a MPS with $\chi_{MPS} \sim 128$. }
\label{fig:MERA_MPS}
\end{figure*}

\subsection{Entanglement in the qMERA circuit}
We can further analyze the accuracy of the qMERA circuit executed on Quantinuum's H1-1 quantum computer by measuring its global entanglement. A useful measure of the amount of entanglement in a state is the bipartite Von-Neumann entanglement entropy $S = -\mathrm{Tr} \rho\log \rho$, where $\rho$ is the reduced density matrix of one half of the system. The entanglement entropy $S$ of a qMERA can be computed directly with a cost that scales exponentially with the qMERA depth, but we found it convenient to  use an MPS approximation to the full qMERA in order to estimate $S$, as shown in Fig.~\ref{fig:MERA_MPS} \cite{miles:2023}. In this way we found a bipartite entanglement entropy of $S_{qMERA}\approx1.593$, which is close to the numerical exact value $S_{exact}\approx1.582$ (obtained by DMRG with high bond dimension). This agreement indicates that the qMERA circuit faithfully captures the long-range entanglement structure of the critical model.

\end{document}